# Dispersion management of a nonlinear amplifying loop mirror-based Erbium-doped fiber laser

**ZBIGNIEW ŁASZCZYCH AND GRZEGORZ SOBOŃ**

*Laser & Fiber Electronics Group, Department of Field Theory, Electronic Circuits and Optoelectronics, Faculty of Electronics, Wrocław University of Science and Technology, Wybrzeże Wyspiańskiego 27, 50-370 Wrocław, Poland*
*\* zbigniew.laszczych@pwr.edu.pl*

**Abstract:** We report an investigation of dispersion management of an all-polarization maintaining Er-fiber oscillator mode-locked via nonlinear amplification loop mirror in a figure-nine cavity configuration with two output ports. The performance of the laser was investigated within the net cavity dispersion ranging from -0.034 $ps^2$ to +0.006 $ps^2$. We show that the spectral and temporal phase of the pulses at both figure-nine outputs have clearly different characteristics. One of the laser outputs provides pulses with significantly better quality; nonetheless, the rejection output also offers ultrashort pulses with broad spectra. Pulses as short as 79 fs with an energy of 83 pJ were generated directly from the laser in the near-zero dispersion regime.

## 1. Introduction

More than three decades of expansion in the field of ultrafast fiber lasers have yielded astonishing development transferred into applications in science [1], industry [2,3], and biomedicine [4,5]. Compact footprint and weight combined with the ability to provide robust and non-interventional operation distinguish fiber laser technology from bulk lasers. Numerous material-based saturable absorbers (SA) such as semiconductor saturable absorber mirrors (SESAM) [6], carbon nanotubes [7], graphene [8], topological insulators [9], transition metal dichalcogenides [10], and black phosphorous [11] have been demonstrated to support reliable and self-starting passive mode-locking. Nevertheless, the dependence of the absorber properties on its design is frequently constraining application to a specific region on the parameter space. A real saturable absorber requires the formulation of a method to introduce it into the laser cavity to achieve proper interaction with light. The demand to establish a reproducible method of manufacturing frequently requires applications of advanced techniques. Moreover, material-based absorbers tend to have a rather low damage threshold in comparison to optical fibers and has an tendency to degrade over time. Therefore, no less attention has been given to the development of additive pulse mode-locking (APM) technology, where the equivalent of the real SA is based on nonlinear effects inside of the laser cavity [12]. In comparison to real SA, such methods highlights their advantages such as ultrafast recovery time, higher damage threshold and the reduction of potential degradation over time [13–17].

The APM methods were introduced more than two decades ago but now experience a growing interest in the ultrafast laser community. They include nonlinear polarization rotation (NPR) [18], nonlinear optical loop mirror (NOLM) [19], nonlinear amplifying loop mirror (NALM) [20], and nonlinear absorbing loop mirror (NAbLM) [21]. The requirement of environmentally stable operation can be fulfilled by the use of all polarization-maintaining (PM) fiber lasers. The NPR method can support all-PM configurations, but it demands additional waveguide design or advanced methods of optimization [22,23]. Both NOLM, NALM also can be made with the use of all-PM fibers and components and are based on a coherent interference between counterpropagating pulses that introduce intensity-dependent loss. The roundtrip transmission of the mirror loop and interference conditions are given by the splitting ratio between both propagation directions. The resultant difference in nonlinear phase shift imitates a fast saturable absorber. In NOLM, the phase shift is an effect of asymmetry of the used coupler. An additional phase difference in NALM configuration is a result of the asymmetry of active fiber position. Self-starting mode-locking without additional non-reciprocal phase bias is achievable but requires high optical pumping to shift transmission curve towards mode-locking supporting part of the slope [24–28].

The phase biasing module can consist of a set of wave plates and one or more Faraday rotators [29,30]. In the so-called "figure of 9" (F9L) configuration, such a phase bias typically consists of two Faraday rotators [31], while setups involving a fiber loop and linear arm might use only one Faraday rotator in the linear part [32,33]. Alternatively, T. Jiang et al. presented an all-PM NALM-based system using reflective-type phase shifter located in the loop that incorporated a Wollaston prism, a Faraday rotator, an $\lambda/8$ waveplate, and a mirror [34]. A recent work by Liu et al. shows a comparison of self-starting threshold and the phase noise between fiber lasers using different phase bias in the linear arm, concluding that usage of $\lambda/8$ waveplate led to the lowest self-starting threshold and highest phase noise level comparing to $\lambda/6$ and $\lambda/10$ waveplates [35]. Kuse et al. reported an all-PM F9L-based stabilized optical frequency comb with phase bias in loop segment incorporating two Faraday rotators and a waveplate resulting in low-noise operation with 40 attoseconds of integrated timing jitter measured from 10 kHz to 10 MHz [31]. An all-PM configuration with a reflection-type shifter, working at center wavelengths of 1030, 1565, and 2050 nm, was successfully commercialized by Menlo Systems GmbH under the trademark figure 9™. Authors reported 3 mW of output power, 3 dB optical spectrum width of 43 nm at a repetition rate of 250 MHz for the setup working at 1565 nm [32]. Incorporation of phase shifter in a reflection segment can reduce the price, scale-up oscillator towards higher repetition rates, and neglect the necessity of use output coupler. The recently demonstrated stable source of 50 fs pulses generated directly from NALM-based fiber oscillator with a pulse energy of 0.16 nJ at 85 MHz incorporated 30:70 fiber coupler and integrated non-reciprocal phase shifter [36]. 44.6 fs pulses from a 257 MHz and 104 mW of output power from mode-locked non-polarization maintaining Er-doped fiber laser based on a biased nonlinear amplifying loop mirror have been reported [37]. Recent publications have demonstrated a robust NALM lasers that were including a 3x3 fiber coupler without additional phase bias, may indicate the necessity of expanding the knowledge in this field [38,39]. Another all-PM NALM-based laser with integrated non-reciprocal phase bias in the loop segment with over 1.1 mW output power and 477 fs pulses at the repetition rate of 121 MHz [40]. Experimental and numerical investigation of mode-locking regimes within wide net cavity dispersion of NALM-based all-PM erbium-doped fiber laser achieved 132 fs pulse with a spectral width of 46 nm in the stretched pulse mode-locking regime. Build-up process dynamics for soliton stretched pulse, and dissipative soliton mode-locking regimes were investigated. Dispersion management provided information about the limits of the performance of the given mode-locking regime. However, no methods capable of opening the impulse phase were used to fully characterize the source and no noise characteristics were measured [41]. Dispersion management of NALM-based fiber laser utilizing a Yb-doped single-polarization large-mode-area photonic crystal fiber working at 1040 nm shown pulses as short as 68 fs (13 nJ) for near-zero dispersion region and 1.95 W (152 fs) of output power in dissipative soliton region, where additional spectral filtering was necessary [42]. A recent investigation of flexible all-PM Yb-doped fiber laser with reflect-type phase biasing in different dispersion-dependent regimes has shown the dependence of intensity noise and free-running linewidth of the carrier-envelope-offset (CEO) on the net cavity dispersion. In combination with far from the spontaneous emission peak of Yb, close to zero net cavity dispersion leads to significant suppression of relative intensity noise (RIN) as well as narrowing of CEO linewidth and indicates the necessity of further investigation of dispersion regimes [43]. A LIDAR system based on two free-running all-PM NALM mode-locked fiber lasers with a ranging accuracy of ±2 μm within 65 m has been demonstrated. The advantage of the presented system was the use of one of the output ports for distance measurement (transmitted), while the reflective port was used for monitoring [44]. A novel simulational and experimental study on steady-state NALM fiber oscillator with both transmitted and reflected output ports has revealed a relationship between the transmission function and fluctuation of the intracavity pulse peak power that leads to amplitude-noise suppression in the transmitted output port [45]. Additionally, NALM has been shown as a mode-locker in various dual-wavelength mode-locked fiber lasers incorporating Sagnac loop filter [46], polarization multiplexing [47], and mechanical spectral filtering [33].

The most comprehensive study on the properties of figure-nine NALM laser so far, including mode-locking build-up dynamics, was performed by Nishizawa et al. [41]. However, none of the referred publications contained an extensive analysis of the spectral and temporal phase of the output pulses, obtained at both output ports as a function of the net cavity dispersion. Expanding knowledge resources in the field of this artificial saturable absorber grants permission to more suitable design solutions of given fiber laser to suit a particular application because the principle of operation of ultrafast optical switching is defined by the temporal characteristic of interfering portions of light. In this work, we report an investigation of dispersion management of all-PM NALM-based Er-fiber oscillator within the net cavity dispersion ranging from -0.034 $ps^2$ to +0.006 $ps^2$, which covers the three fundamental mode-locking regimes (soliton, stretched-pulse, and dissipative soliton). Our laser incorporates a reflection-type nonreciprocal phase shifter based on only one Faraday rotator, which reduces the number of components and simplifies the entire system in comparison to [29–31,37,41,42,47,48]. Flexibility and ease in the selection of saturable absorber properties is a unique feature of the presented setup. We show that the pulses generated from both outputs of the figure-nine laser have different shapes of the spectral and temporal phase. One of the laser outputs provides pulses with a clearly better quality. Comprehensive spectral and temporal characterization, including phase analysis of the presented pulsed source, broadens the knowledge on the operation of ultrafast lasers using NALM as a saturable absorber.

**2. Experimental setup and results**

The structure of the oscillator consisting of a NALM segment and a linear arm is shown in Fig. 1. The fiber part includes 94.5 cm of polarization-maintaining (PM) erbium-doped fiber (Liekki Er80-4/125-PM, EDF), PM single-mode fiber, and a PM wavelength division multiplexer. The gain fiber is placed asymmetrically in the loop, and it is pumped by a pigtailed single-mode laser diode operating at 980 nm (3SP Technologies, 1999CVB, LD). Counterpropagating in the loop pulses are perpendicularly coupled into the linear arm and combined on a polarizing beamsplitter cube (Thorlabs PBS124, $PBC_1$). The non-reciprocal phase shifter consisting of a quarter-wave plate (Thorlabs WPQ05M-1550, $QWP_1$) and a Faraday rotator (Thorlabs I1550R5, FR) introduces phase bias between counter-circulating portions of light. A fast saturable absorber is mitigated at $PBC_2$ when coherent interference between counterpropagating pulses introduce intensity-dependent loss. The rejected part of the light is coupled out to the output 2 when the transmitted part passes through $QWP_2$ and is reflected by the rear mirror (Thorlabs PF10-03-P01, M). The splitting ratio at output port 1 and modulation depth can be adjusted by fine-tuning the angle of $QWP_2$, while the linear losses and splitting ratio at output port 2 can be adjusted by a change of $QWP_1$ angle.

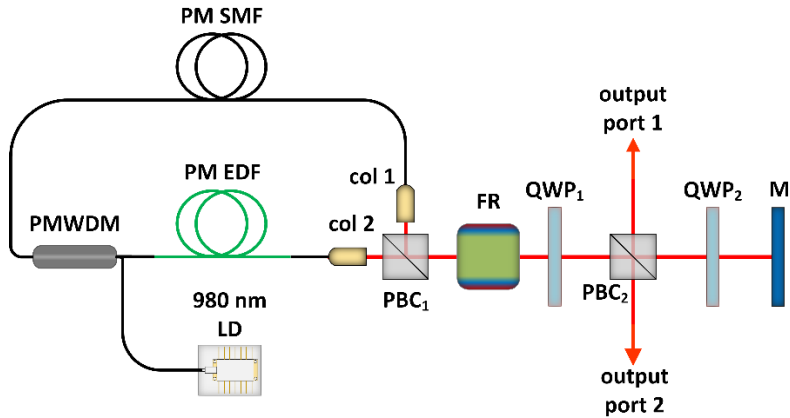

Fig. 1. Experimental setup. PM WDM: polarization maintaining wavelength division multiplexer; PM EDF: polarization-maintaining erbium-doped fiber; PM SMF: polarization-maintaining single-mode fiber; LD: laser diode; col: collimator; PBC; polarization beam combiner; FR: Faraday rotator; QWP: quarter-wave plate; M: mirror.

We started our investigation of the laser performance with the PM SMF length of 290 cm, resulting in a net cavity dispersion of -0.034 ps$^2$. Once the appropriate angular position of QWP$_1$ and QWP$_2$ was found, self-starting mode-locking could be achieved by increasing the pump power of LD to the level of 355 mW. Initially, the multi-soliton state was observed. Afterward, the single-pulse operation was obtained by decreasing the LD power to 128 mW. Output characteristics of each output port have been recorded via an optical spectrum analyzer (Yokogawa AQ6376, OSA), a 3.6 GHz RF spectrum analyzer (Agilent EXA N9010A, RF), an autocorrelator (APE pulseCheck, AC), and second harmonic frequency-resolved optical gating technique system (Mesa Photonics FS-Ultra2, FROG). The results are shown in Fig. 2. The full width at half maximum (FWHM) of the optical spectrum was 14 nm and 26 nm for output ports 1 and 2, respectively. Despite the solitonic character of pulses, the appearance of Kelly sidebands is limited what can be explained by the relatively low value of the net cavity dispersion and spectral filtering in NALM. The RF spectrum measurement indicated a repetition frequency of 51.55 MHz, with a signal to noise ratio in the RF signal of over 65 dB Autocorrelation traces indicated pulse duration equal to 250 fs and 243 fs with an average output power of 3.5 mW and 5.0 mW for output ports 1 and 2, respectively. Reconstructed temporal intensities imply the occurrence of a slight temporal chirp of pulses at output port 2. In contrast, the temporal phase remains flat at output port 1 and indicates minor differences in pulse duration concerning autocorrelation measurements. Calculated trace-area-normalized FROG error was below 1% for both output ports.

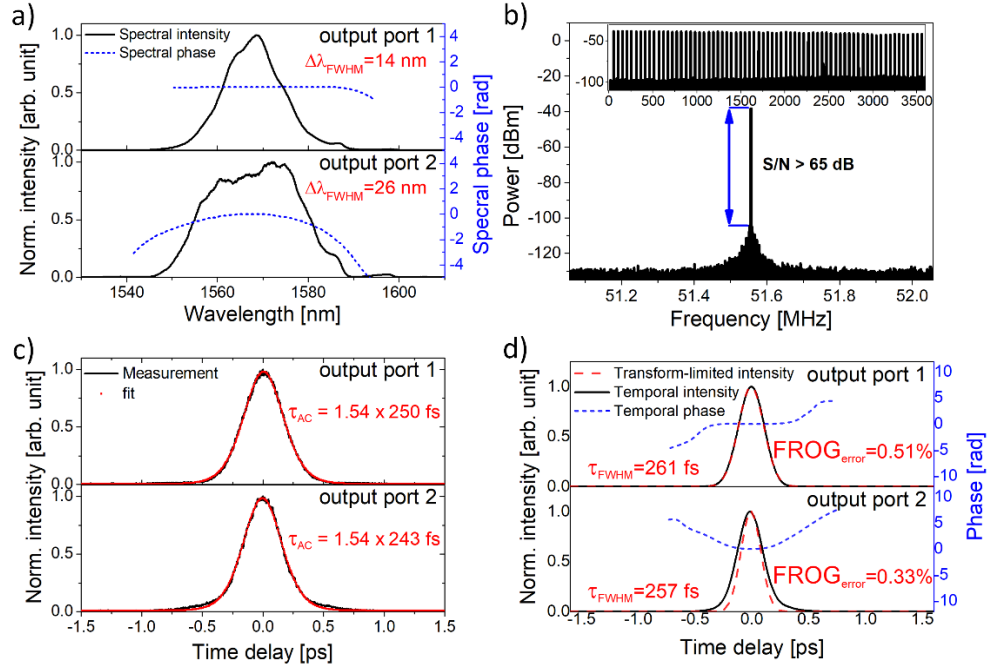

Fig. 2. Measured output characteristics of both NALM output ports with net cavity dispersion equal to -0.034 ps$^2$. (a) Optical spectra and spectral phases measured at both output ports. (b) Output port 1 RF spectrum (the resolution bandwidth of the measurement was 51 Hz); inset, RF spectrum measured in a 3.5 GHz span. (c) Autocorrelation traces of both output ports (d) Reconstructed temporal intensity (black line), transform-limited intensity (dashed red line), and temporal phase of the pulse (dashed blue line) for both output ports.

Once stable and repeatable self-starting laser operation was attained, we investigated the laser operation for different values of net cavity dispersions, realized by reduction of PM SMF length in the NALM. Compared to other works [41,49,50], we were able to establish net cavity dispersion ranging from soliton regime to net-normal dispersion regime by only appropriate balance between active (normal dispersion) and passive (anomalous dispersion) fiber, without the need of adding additional dispersion compensation fiber segments. To determine the net cavity dispersion, we assumed the group velocity dispersion (GVD) as 28.0 ps$^2$/km and -23.0 ps$^2$/km at 1550 nm for active and passive fiber, respectively. Within the net anomalous dispersion region, the laser works in the soliton-like regime, and optical spectra show the typical shape for fundamental soliton pulses with minor Kelly's sidebands (-0.019 ps$^2$ and -0.012 ps$^2$ of Fig. 3(a) and Fig. 4(a)). The self-starting operation was achievable but required slightly higher pumping power in a range of 355 to 415 mW, while single pulse operation in this dispersion regime was occurring for pump powers close to 125 mW. Successive reduction of the passive fiber length led to a shift into stretched pulse regime, indicated by broadening of the optical spectra, decrement of pulse duration, and increase of output power at both output ports (-0.001 ps$^2$ of Fig. 3 and Fig. 4). Self-starting in this region was achieved without multi-pulse operation. Instead of that, a single pulse with a continuous wave (CW) component appeared for the pump power around 530 mW. Further reduction of pump power was eliminating this feature. Stable, single pulse operation in near-zero dispersion for a pumping power of 180 mW and a repetition frequency of 80.12 MHz has reached pulses as short as 79 fs and 84 fs for output 1 and output 2, respectively (-0.001 ps$^2$ of Fig. 3(b) and Fig. 4(b)). In the net-normal dispersion regime, we observed further spectral broadening while the autocorrelation traces indicated slightly extended pulses durations. Pumping power as high as 555 mW was needed to obtain self-starting operation with CW component, while single pulse operation was reachable for over 240 mW (+0.001 ps$^2$ and +0.003 ps$^2$ of Fig. 3(a) and Fig. 4(a)). No additional spectral filtering was necessary to observe dissipative solitons. A possible explanation of this feature is the spectral limitation resulting from the active fiber gain bandwidth, also assumed by Nishizawa et al. in their numerical investigation of pulse dynamics in this dispersion region [41]. When moving further towards the net-normal dispersion regime, the spectral width has narrowed, and the temporal width was increased to above 500 fs for both output ports (+0.006 ps$^2$ of Fig. 3(b) and Fig. 4(b)). The self-starting operation with CW component occurred for pumping over 530 mW, and decreasing it to the level of 450 mW was sufficient to access single pulse operation. For the net cavity dispersion equal +0.006 ps$^2$, we obtained the highest values of average output power equal to 16 mW and 37 mW for output 1 and 2, respectively.

Optical spectra and corresponding autocorrelation traces directly from the oscillator are shown in Fig. 3. for output port 1, and in Fig. 4. for output 2. Asymmetry in the position of active fiber leads to a situation where the pulse entering the fiber loop via collimator 2 is amplified to high peak power and is broadened through self-phase modulation during propagation in a long segment of PM SMF. On the other hand, the pulse entering the loop via collimator 1 is amplified in the active fiber just before leaving the loop. Dudley et al. suggested that even for optimal ultrafast optical switching, the interference of two pulses with different duration may lead to a complex reflected pulse at the reflect port, which in our case is output port 2 [51]. Autocorrelation traces recorded for net cavity dispersion -0.019 ps$^2$ and -0.012 ps$^2$ presents typical for soliton-like regime shapes given by sech$^2$. Once the near-zero dispersion regime occurs, pulses can be approximated by Gaussian shape. Pulses generated from both outputs had a duration below 100 fs, and the shortest pulse duration of 79 fs with an average power of 6.6 mW was achieved for net cavity dispersion -0.001 ps$^2$ at output port 1. The time-bandwidth products were equal to 0.605 and 0.717 at output ports 1 and 2, respectively. Further increment of net dispersion leads to an increment of the time-bandwidth product (TBP). Moreover, rectangle-like spectra for an all-normal dispersion regime are typical for dissipative soliton operation [52].

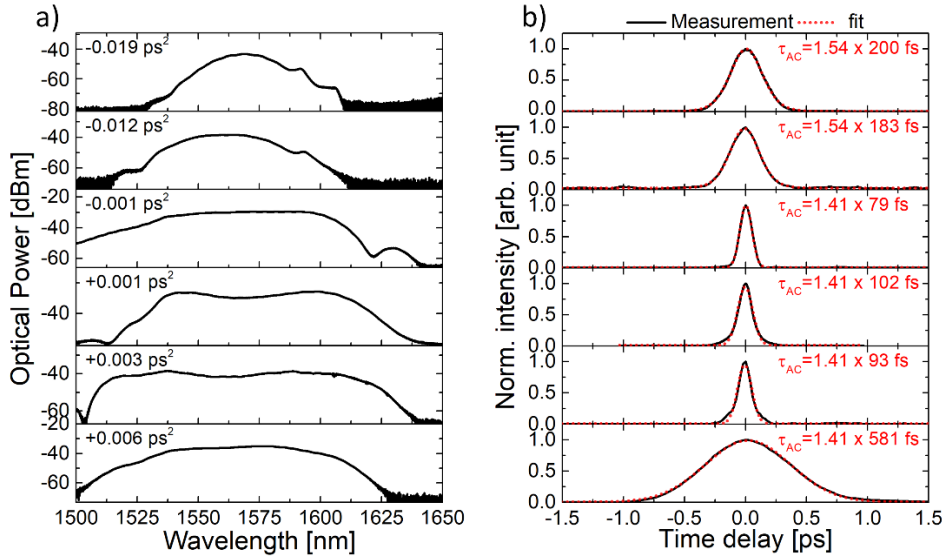

Fig. 3. Optical spectra (a) and corresponding autocorrelation traces of the pulses (b) generated on output port 1 as the function of the net cavity dispersion.

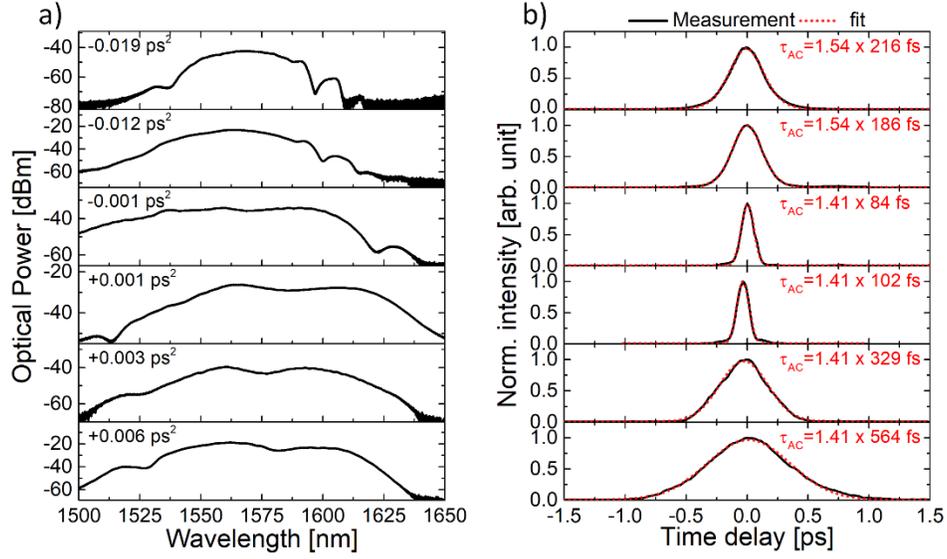

Fig. 4. Optical spectra (a) and corresponding autocorrelation traces of the pulses (b) generated on output port 2 as the function of the net cavity dispersion.

The results of detailed spectral and temporal phase analysis of both outputs at different net dispersions are shown in Figures 5 (output 1) and 6 (output 2). The FROG-derived electric fields measured for the solitonic and near-zero regimes indicated pulses with flat, both spectral and temporal phase at output 1. The result confirms the results obtained by the autocorrelation measurement, thus ensuring the appropriate utility of the presented configuration to the generation of a stable train of ultrashort pulses. Once the net dispersion takes a positive value, the spectral phase indicates a spectral chirp, and the spectrum is red-shifted. Still, with further shortening of the cavity, it shifts towards shorter wavelengths. Considering the dissipative soliton-like character of the pulses for the net cavity dispersion equal to +0.006 ps$^2$, the measured spectral and temporal phase indicates rather usual chirp under the absence of sufficiently large anomalous dispersion [52]. Despite the increase in FWHM value, the temporal duration is increasing, confirming the deviation from the TBP limit in this regime.

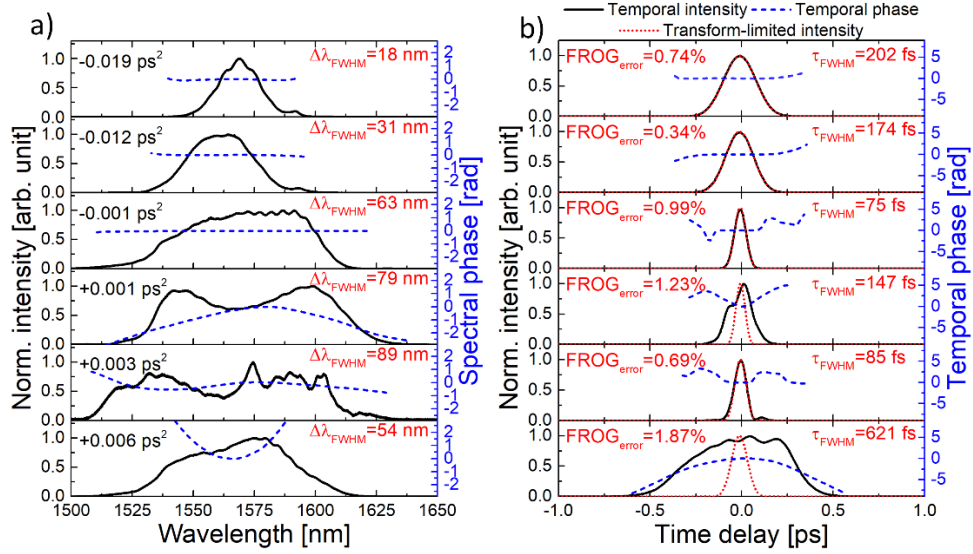

Fig. 5. (a) Optical spectra (solid black line) and spectral phase (dashed blue line) (b) reconstructed temporal intensity (solid black line), transform-limited intensity (dotted red line), and temporal phase (dashed blue line) measured at the output port 1.

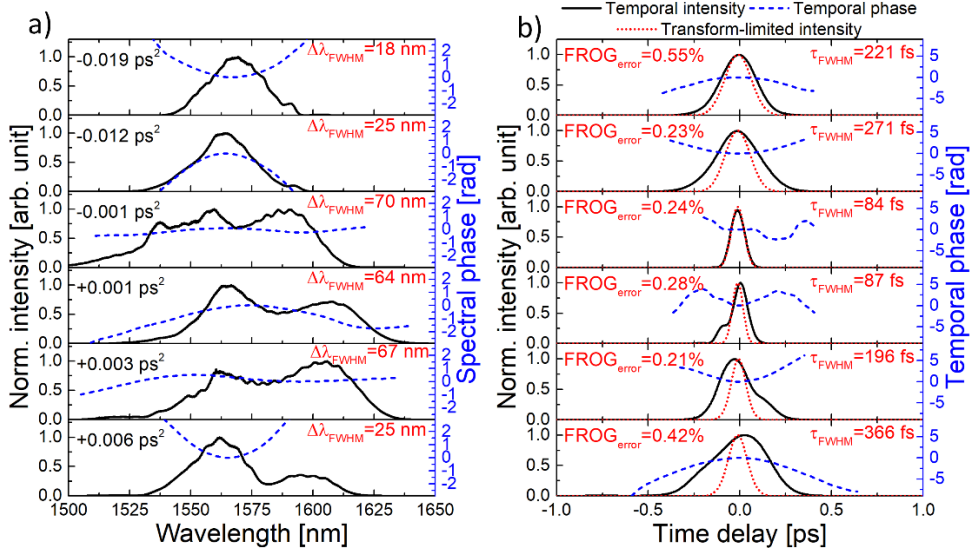

Fig. 6. (a) Optical spectra (solid black line) and spectral phase (dashed blue line) (b) reconstructed temporal intensity (solid black line), transform-limited intensity (dotted red line), and temporal phase (dashed blue line) measured at the output port 2.

FROG measurement for output port 2 is shown in Fig. 6. FROG-derived electric fields measured for the solitonic regime indicate slightly chirped pulses. Gain asymmetry of the NALM segment manifests in differences in temporal shapes of counterpropagating pulses affecting the shape of pulses at the rejection port [51]. Interestingly, in the near-zero dispersion region, no spectral chirp or distortion is observed. In the stretched-pulse operation regime, the contrast in dispersion value in different segments of the cavity disturbs phase matching of dispersive waves, leading to greater nonlinear phase accumulation per roundtrip. Since the counterpropagating pulses experience different phase shifts, then their temporal and spectral shapes are contrasting with each other when interference occurs. Further shortening of the SMF segment leads to an increase of asymmetry in active fiber position. Therefore, the difference in the optical spectrum between outputs might be observed.

Figure 7 summarizes the pulse spectral widths, durations, and energies as the function of the net dispersion of the cavity, recorded at both outputs. Anomalous dispersion region limits pulse width and energy what can be explained by the soliton area theorem. The maximum average power is here limited by the single-pulse operation limit. The decrease in pulse duration and increment of pulse energy can be observed in the near-zero dispersion regime, where the stretched-pulse mechanism leads to periodical stretching and recompressing allows for the accumulation of energy in the pulse. Dissipative soliton operation in a positive net dispersion region contributes to the increase of TBP and further growth of the pulse energy. Both outputs show the capacity to utilize them in spite of their slightly different output characteristics. Notably, output port 2 reached the output power of 37 mW (over 400 pJ of energy per pulse) for pulse duration of 564 fs.

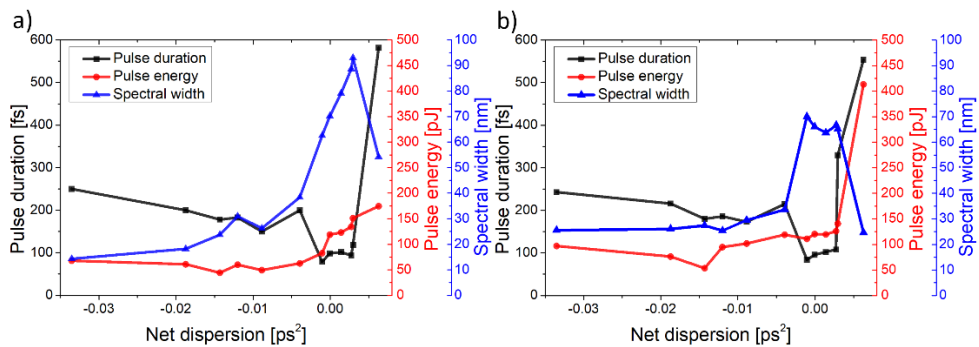

Fig. 7. Pulse duration, pulse energy, and spectral width (FWHM) as a function of net cavity dispersion for (a) output port 1; (b) output port 2.

## 3. Summary and conclusions

Summarizing, we have demonstrated an investigation of dispersion management of all-PM NALM-based Er-fiber oscillator within the net cavity dispersion ranging from -0.034 $ps^2$ to +0.006 $ps^2$. Pulses short as 79 fs generated directly from the laser with a pulse energy of 83 pJ for near-zero dispersion regime. Dispersion management is an enabling method in the way of optimization of laser operation for further applications. Output characteristics of both ports differ in terms of pulse duration, energy, and spectral bandwidth, but, also in terms of phase. We presented the first, to our best knowledge, demonstration of retrieved spectral and temporal phases of the pulses obtained at two outputs of a NALM-based ultrafast fiber laser. One of the laser outputs provides pulses with significantly better quality (i.e., flat spectral phase), especially in the anomalous dispersion regime. Nevertheless, we have experimentally shown that at near-zero net cavity dispersion both outputs are capable of delivering unchirped, sub-80 fs pulses.

## Funding